\documentclass[aps,prb,reprint,superscriptaddress,amsmath,amssymb]{revtex4-2}

\usepackage{graphicx}
\usepackage{bm}
\usepackage{hyperref}
\hypersetup{hidelinks}

\begin{document}

\title{Altermagnetism across the BCS-BEC crossover}

\author{Iogann Tolbatov}
\email[Author to whom correspondence should be addressed: ]{itolbatov@uniss.it}
\affiliation{Department of Chemical, Physical, Mathematical and Natural Sciences, University of Sassari, 07100 Sassari, Italy}

\author{Luca Salasnich}
\affiliation{Dipartimento di Fisica e Astronomia ``Galileo Galilei'' and Padua QTech Center, Universit\`a di Padova, Via Marzolo 8, 35131 Padova, Italy}
\affiliation{INFN Sezione di Padova, Via Marzolo 8, 35131 Padova, Italy}

\begin{abstract}
We study a two-dimensional paired Fermi system in which altermagnetism produces a spin splitting that depends on momentum. The interaction strength is described through the two-body binding energy, so that the chemical potential and pairing gap evolve self-consistently from the weak-coupling Bardeen-Cooper-Schrieffer (BCS) regime to the strong-coupling Bose-Einstein-condensate (BEC) regime at fixed density. The stability of the uniform paired state is examined by giving the pairs a small center-of-mass momentum and following the resulting change in free energy. We show that at zero temperature, the phase stiffness follows a universal quadratic suppression, $J/J_0 = 1 - (\alpha/\alpha_0)^2$, across the entire fully gapped crossover regime. Deviations from this relation emerge only on the BCS side upon the opening of gapless Bogoliubov pockets, which rapidly reduce the stiffness and can trigger an instability toward finite-momentum pairing. In the BEC regime, this universal quadratic correction corresponds to the altered effective mass of the tightly bound composite bosons. The model therefore provides a simple setting in which to compare altermagnetic pair breaking on the BCS and BEC sides of the crossover.
\end{abstract}

\maketitle

\section{Introduction}

Altermagnets are collinear magnetic systems with zero net magnetization but spin-split electronic bands, originating from their underlying microscopic multipole structure \cite{hayami2019, hayami2026}, whose splitting depends strongly on crystal momentum \cite{smejkal2022beyond,smejkal2022landscape}. For the continuum problem considered here, a useful physical picture is that altermagnetism acts like a Zeeman field whose strength and sign depend on momentum. The two spin species are therefore shifted in opposite ways, but the mismatch is not the same in every direction in momentum space.

This directional spin splitting is relevant to superconductivity because conventional spin-singlet pairing combines fermions of opposite spin and opposite momentum. A mismatch between the two partners can weaken uniform pairing. In ordinary spin-imbalanced superconductors, the same basic competition can lead to Fulde-Ferrell-Larkin-Ovchinnikov (FFLO) states, in which the pairs acquire a finite center-of-mass momentum \cite{fulde1964,larkin1965,casalbuoni2004}. Recent studies have shown that altermagnetic band structures can also favor finite-momentum superconductivity and other unconventional superconducting responses \cite{chakraborty2024finiteq,banerjee2024diode,fu2025,fu2026,cayao2025}. To date, most theoretical explorations of these states rely on numerical Bogoliubov–de Gennes lattice calculations, real-space phase-twist implementations, or quasiclassical Usadel frameworks near $T_c$ \cite{liu2026altermagnetic,hong2025,delasheras2025}. A notable recent exception is the analytical study by Liu, Hu, and Liu \cite{liu2026fflo}, which examines a two-dimensional continuum model—equivalent to the one considered here upon a $\pi/4$ rotation of the momentum axes. By analyzing this model at fixed particle number, they analytically map out a ground-state phase diagram featuring topological Bogoliubov Fermi surfaces and FFLO states. In complementary lattice model studies, the stability of uniform pairing has also been examined by evaluating the phase stiffness at low filling \cite{liu2026altermagnetic}.

The BCS-BEC crossover offers a complementary way to organize the problem and builds upon this analytical picture. In two dimensions, an attractive contact interaction always supports a two-body bound state, and the binding energy provides a convenient physical measure of the interaction strength. At fixed density, increasing this attraction drives the chemical potential from positive values, characteristic of overlapping Cooper pairs, to negative values, characteristic of tightly bound composite bosons \cite{randeria1989,randeria1990}. While existing continuum and dilute-lattice treatments of altermagnetic superconductors focus predominantly on the weakly interacting metallic regime with a well-defined Fermi surface ($\mu > 0$) \cite{liu2026altermagnetic,hong2025,liu2026fflo}, our work establishes what happens when the pairing strength is tuned into the non-metallic regime. By explicitly calculating the phase stiffness across the entire crossover, we extend previous analyses to show how tightly bound composite bosons shield the directional spin mismatch. The crossover therefore lets us ask a simple question: does the altermagnetic mismatch remain equally effective once the two fermions are bound into a compact pair?

We address this question with a minimal continuum model. We do not attempt a complete microscopic classification of all pairing channels allowed in altermagnets, because those constraints depend on the detailed orbital and sublattice structure \cite{chakraborty2025constraints}. Instead, we assume an effective spin-singlet, local attractive interaction and focus on two quantities that have a direct physical meaning across the crossover: the self-consistent pairing state at fixed density and the energy cost of giving the pairs a small center-of-mass momentum.

It is instructive to contrast this altermagnetic pairing mechanism with the effects of synthetic spin-orbit coupling (SOC) in ultracold Fermi gases \cite{Wang2014, Cheuk2012, Zhai2015}. Both phenomena introduce a momentum-dependent spin splitting that can distort the Fermi surfaces and drive FFLO-type pairing instabilities. However, there are fundamental differences between the two. Artificial SOC inherently breaks spatial inversion symmetry and strongly induces singlet-triplet mixing in the pairing channel, heavily modifying the BCS-BEC crossover \cite{DellAnna2011, DellAnna2012}. In contrast, the continuum altermagnetic model breaks time-reversal and combined $\mathcal{PT}$ symmetries, but preserves spatial inversion symmetry ($\mathcal{P}$), producing a $d$-wave-like Zeeman splitting. Furthermore, while the altermagnetic spin splitting can induce odd-frequency triplet pair correlations \cite{maeda2025}, we assume a spin-singlet attractive interaction and order parameter. This isolates the specific thermodynamic effects of momentum-dependent pair breaking on the singlet condensate, avoiding the added complexity of mixed-parity pairing.

The organization of the calculation mirrors this physical picture. While microscopic path-integral derivations of effective phase actions exist for related broken-symmetry phases like charge-density waves~\cite{yang2026}, a direct analytical continuum approach bridging the BCS and BEC regimes in altermagnets provides a transparent formulation of the pair stiffness. We first introduce the normal-state dispersion and the Bogoliubov spectrum. We then determine the chemical potential and gap from the regularized two-dimensional gap and number equations. Finally, we allow the pairs to move and extract the phase stiffness from the curvature of the free energy. This route is sufficient to distinguish a locally stable uniform state from an instability toward finite-momentum pairing, while avoiding intermediate kernels that do not enter the final observables.

\section{Continuum model and self-consistent pairing}

We consider a two-dimensional gas of fermionic particles with mass $m$, which can be physically realized in thin-film altermagnetic heterostructures (such as candidate RuO$_2$-based heterostructures \cite{smejkal2022beyond,smejkal2022landscape,choi2026exploring,jeong2026altermagnetic,lee2026strain}) or two-dimensional electron gases with proximity-induced pairing \cite{shabani2016two,kjaergaard2016quantized}.

The single-particle states of these fermions are characterized by the two-dimensional crystal momentum vector $\mathbf{k} = (k_x, k_y)$, with magnitude $k=|\mathbf k|$. The isotropic kinetic energy measured from the chemical potential $\mu$ is
\begin{equation}
\xi_{\mathbf k}=\frac{\hbar^2 k^2}{2m}-\mu,
\label{eq:xi}
\end{equation}
where $\hbar$ is the reduced Planck constant.

The defining feature of the altermagnetic state is a momentum-dependent spin splitting that distorts the Fermi surfaces of opposite spins in orthogonal directions. We model this term as an effective, $d$-wave Zeeman-like field \cite{smejkal2022beyond}
\begin{equation}
J_{\mathbf k}=\alpha(k_x^2-k_y^2).
\label{eq:Jk}
\end{equation}

The parameter $\alpha$ quantifies the strength of this altermagnetic splitting. Experimentally, the magnitude of $\alpha$ is proposed to be tunable via epitaxial strain, chemical doping, or gate voltages in thin-film geometries \cite{khodas2026tuning,zhang2026tunable,wang2026strain,lee2026strain}.

It has the same dimensions as $\hbar^2/(2m)$. For later convenience we define the natural continuum scale
\begin{equation}
\alpha_0=\frac{\hbar^2}{2m}.
\label{eq:alpha0}
\end{equation}
The continuum spectrum remains bounded from below only when
\begin{equation}
0\leq \alpha<\alpha_0.
\label{eq:alphabound}
\end{equation}
A microscopic lattice motivation for Eq.~\eqref{eq:Jk} is summarized in Appendix~\ref{app:lattice}.

The spin-up dispersion is
\begin{equation}
\xi_{\mathbf k\uparrow}=\xi_{\mathbf k}+J_{\mathbf k}.
\label{eq:updisp}
\end{equation}
The spin-down dispersion is
\begin{equation}
\xi_{\mathbf k\downarrow}=\xi_{\mathbf k}-J_{\mathbf k}.
\label{eq:downdisp}
\end{equation}
Thus the two spin species have opposite distortions of their constant-energy contours. This is the main normal-state ingredient that distinguishes the present problem from the usual isotropic BCS-BEC crossover.

We now add an effective local attraction of strength $g>0$ in the spin-singlet channel. In two dimensions the bare coupling $g$ is ultraviolet divergent and is replaced by the physical two-body binding energy $\epsilon_B$. We denote the free-particle kinetic energy by
\begin{equation}
\epsilon_{\mathbf k}=\frac{\hbar^2k^2}{2m}.
\label{eq:epsilonk}
\end{equation}
The regularization relation is
\begin{equation}
\frac{1}{g}=\int\frac{d^2\mathbf k}{(2\pi)^2}\frac{1}{2\epsilon_{\mathbf k}+\epsilon_B}.
\label{eq:regularization}
\end{equation}
Here $\epsilon_B>0$ is the binding energy of the two-body bound state in vacuum.

Let $\Delta_0$ be the real and spatially uniform pairing gap. In the absence of the altermagnetic shift, the usual Bogoliubov energy is
\begin{equation}
E_{\mathbf k}=\sqrt{\xi_{\mathbf k}^2+\Delta_0^2}.
\label{eq:Ek}
\end{equation}
One quasiparticle branch is
\begin{equation}
E_{\mathbf k}^{+}=E_{\mathbf k}+J_{\mathbf k}.
\label{eq:Eplus}
\end{equation}
The other quasiparticle branch is
\begin{equation}
E_{\mathbf k}^{-}=E_{\mathbf k}-J_{\mathbf k}.
\label{eq:Eminus}
\end{equation}
Because $J_{\mathbf k}$ changes sign with direction, either branch can approach zero in selected regions of momentum space. Gapless Bogoliubov pockets appear when the magnitude of the altermagnetic shift exceeds the positive Bogoliubov energy.

At fixed total areal density $n$, the two unknown mean-field quantities are $\mu$ and $\Delta_0$. They must be found together. The regularized gap equation is
\begin{equation}
0=\int\frac{d^2\mathbf k}{(2\pi)^2}
\left[
\frac{1}{2\epsilon_{\mathbf k}+\epsilon_B}
-\frac{1-f(E_{\mathbf k}^{+})-f(E_{\mathbf k}^{-})}{2E_{\mathbf k}}
\right].
\label{eq:gap}
\end{equation}
The number equation is
\begin{equation}
n=\int\frac{d^2\mathbf k}{(2\pi)^2}
\left[
1-\frac{\xi_{\mathbf k}}{E_{\mathbf k}}
\left(1-f(E_{\mathbf k}^{+})-f(E_{\mathbf k}^{-})\right)
\right].
\label{eq:number}
\end{equation}
The function $f(E)$ is the Fermi-Dirac distribution. The temperature is denoted by $T$, Boltzmann's constant by $k_B$, and $\beta=1/(k_BT)$, so that $f(E)=[\exp(\beta E)+1]^{-1}$.

\begin{figure*}[t]
\centering
\includegraphics[width=\textwidth]{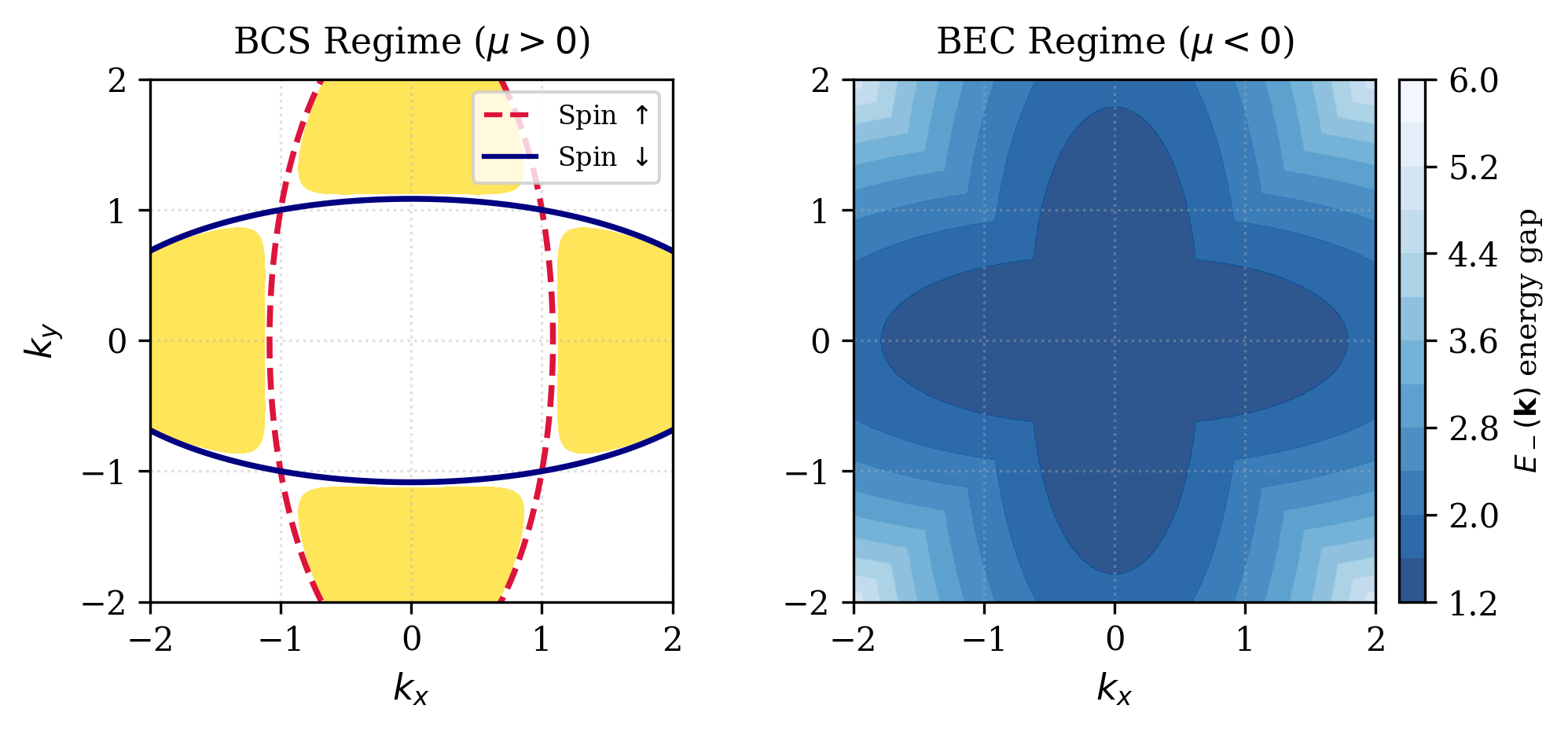}
\caption{Schematic view of the quasiparticle structure across the BCS-BEC crossover. Energies are measured in units of the Fermi energy $\epsilon_F$ and momenta in units of the density-defined reference momentum $k_F = \sqrt{4\pi n}$. The right panel displays the lower Bogoliubov quasiparticle branch, $E_{\mathbf{k}} - |J_{\mathbf{k}}|$, explicitly showing the excitation gap. In the BCS regime (left panel, calculated for $\mu/\epsilon_F=1.0, \alpha/\alpha_0=0.35, \Delta/\epsilon_F=0.25$), the momentum-dependent spin splitting deforms the two spin Fermi surfaces in orthogonal directions and creates gapless Bogoliubov pockets where this lower branch becomes negative. In the BEC regime (right panel, calculated for $\mu/\epsilon_F=-1.0, \alpha/\alpha_0=0.35, \Delta/\epsilon_F=0.8$), the fermionic quasiparticle spectrum remains fully gapped. The spectrum can still be anisotropic: the weaker sensitivity of tightly bound pairs to altermagnetism is instead a property of their center-of-mass motion, discussed below.}
\label{fig:fermi_topology}
\end{figure*}

It is useful to express the density through the Fermi energy of the noninteracting two-dimensional gas,
\begin{equation}
\epsilon_F=\frac{\pi\hbar^2n}{m}.
\label{eq:epsilonF}
\end{equation}
At zero temperature, if both quasiparticle branches stay positive, the occupation factors vanish. The gap and number equations then reduce to the standard two-dimensional mean-field relations \cite{randeria1989,randeria1990}. The chemical potential is
\begin{equation}
\mu=\epsilon_F-\frac{\epsilon_B}{2}.
\label{eq:mu2d}
\end{equation}
The pairing gap is
\begin{equation}
\Delta_0=\sqrt{2\epsilon_F\epsilon_B}.
\label{eq:gap2d}
\end{equation}
These simple expressions cease to apply once gapless pockets are present, because the zero-temperature occupation factors are then nonzero in part of momentum space. In that regime Eqs.~\eqref{eq:gap} and \eqref{eq:number} must be solved with the altermagnetic splitting included explicitly, altering the Fermi surface topology as illustrated in Fig.~\ref{fig:fermi_topology}.

\section{Finite pair momentum and phase stiffness}

The phase stiffness measures how much energy is required to make the condensate phase vary slowly in space. A particularly transparent way to calculate it is to give every pair the same small center-of-mass momentum. We therefore write the order parameter as
\begin{equation}
\Delta(\mathbf r)=\Delta\,e^{2i\mathbf q\cdot\mathbf r}.
\label{eq:phase_twist}
\end{equation}
Here $\mathbf r$ is the two-dimensional position, $\Delta$ is the magnitude of the pairing field for the state under consideration, $i$ is the imaginary unit, and $\mathbf q=(q_x,q_y)$ is the phase-gradient wavevector. The physical center-of-mass momentum of a pair is $2\hbar\mathbf q$.

For finite $\mathbf q$, an up-spin fermion at momentum $\mathbf k+\mathbf q$ pairs with a down-spin fermion at momentum $-\mathbf k+\mathbf q$. The average energy of the two partners is
\begin{equation}
\bar\xi_{\mathbf k\mathbf q}=\frac{\xi_{\mathbf k+\mathbf q,\uparrow}+\xi_{-\mathbf k+\mathbf q,\downarrow}}{2}.
\label{eq:xibar}
\end{equation}
Their mismatch is
\begin{equation}
\zeta_{\mathbf k\mathbf q}=\frac{\xi_{\mathbf k+\mathbf q,\uparrow}-\xi_{-\mathbf k+\mathbf q,\downarrow}}{2}.
\label{eq:zeta}
\end{equation}
The corresponding positive Bogoliubov energy is
\begin{equation}
E_{\mathbf k\mathbf q}=\sqrt{\bar\xi_{\mathbf k\mathbf q}^{\,2}+\Delta^2}.
\label{eq:Ekq}
\end{equation}
The first finite-momentum quasiparticle branch is
\begin{equation}
E_{\mathbf k\mathbf q}^{(+)}=E_{\mathbf k\mathbf q}+\zeta_{\mathbf k\mathbf q}.
\label{eq:Ekqplus}
\end{equation}
The second finite-momentum quasiparticle branch is
\begin{equation}
E_{\mathbf k\mathbf q}^{(-)}=E_{\mathbf k\mathbf q}-\zeta_{\mathbf k\mathbf q}.
\label{eq:Ekqminus}
\end{equation}
At zero pair momentum, the average energy reduces to $\xi_{\mathbf k}$ and the mismatch reduces to $J_{\mathbf k}$. The uniform spectrum of the previous section is therefore recovered continuously.

The mean-field grand-potential density for a state with pair wavevector $\mathbf q$, gap magnitude $\Delta$, and chemical potential $\mu$ is
\begin{equation}
\begin{split}
\omega(\mathbf q;\Delta,\mu)=&\frac{\Delta^2}{g}
+\int\frac{d^2\mathbf k}{(2\pi)^2}\left(\bar\xi_{\mathbf k\mathbf q}-E_{\mathbf k\mathbf q}\right)\\
&-\frac{1}{\beta}\int\frac{d^2\mathbf k}{(2\pi)^2}
\sum_{s=\pm1}\ln\!\left[1+e^{-\beta(E_{\mathbf k\mathbf q}+s\zeta_{\mathbf k\mathbf q})}\right].
\end{split}
\label{eq:omegaq}
\end{equation}
The index $s$ takes the two values $+1$ and $-1$ and simply collects the two quasiparticle branches in a compact notation.

At fixed particle density, the chemical potential $\mu$ acts as a Lagrange multiplier and therefore readjusts when a finite pair wavevector $\mathbf{q}$ is imposed; we denote the corresponding self-consistent value by $\mu_{\mathbf{q}}$. Similarly, $\Delta_{\mathbf{q}}$ denotes the self-consistent gap magnitude. The corresponding Helmholtz free-energy density is
\begin{equation}
\mathcal F(\mathbf q)=\omega(\mathbf q;\Delta_{\mathbf q},\mu_{\mathbf q})+\mu_{\mathbf q}n.
\label{eq:Fq}
\end{equation}
The quantities $\Delta_{\mathbf q}$ and $\mu_{\mathbf q}$ are chosen so that the pairing condition and fixed-density condition remain satisfied for that value of $\mathbf q$. At $\mathbf q=0$ they reduce to the uniform values $\Delta_0$ and $\mu$ obtained from Eqs.~\eqref{eq:gap} and \eqref{eq:number}.

The phase stiffness tensor is the curvature of this self-consistent free energy around the uniform state,
\begin{equation}
J_{ij}=\frac{1}{2}
\left.\frac{\partial^2\mathcal F(\mathbf q)}{\partial q_i\partial q_j}\right|_{\mathbf q=0}.
\label{eq:stiffness}
\end{equation}
The indices $i$ and $j$ denote the Cartesian directions $x$ and $y$. The factor one half follows from the phase convention in Eq.~\eqref{eq:phase_twist}. Because the full momentum dependence is retained before differentiation, the stiffness automatically contains the derivatives of both the ordinary kinetic energy and the altermagnetic splitting.

To connect this definition to the standard macroscopic phase-gradient free energy $\delta \mathcal{F} = \frac{1}{2}\rho_s |\nabla\theta|^2$, we note that a uniform pair momentum $\mathbf{q}$ corresponds to a spatially varying phase $\theta(\mathbf{r}) = 2\mathbf{q}\cdot\mathbf{r}$. Substituting this into the gradient energy yields $\delta \mathcal{F} = 2\rho_s q^2$. Comparing this with our free energy curvature implies that our stiffness $J$ is related to the conventional superfluid stiffness by $\rho_s = J/2$.

For normalization we use the zero-temperature stiffness of the isotropic continuum gas,
\begin{equation}
J_0=\frac{\hbar^2n}{2m}.
\label{eq:J0}
\end{equation}
The symmetry that exchanges the $x$ and $y$ directions together with the two spin species implies that the uniform state has equal diagonal stiffnesses. A positive stiffness means that the uniform state is a local minimum of the free energy. If the stiffness becomes negative, the uniform state is locally unstable and the minimum must move to a finite pair momentum. This identifies an FFLO-type instability, but it does not by itself determine whether the final state is of Fulde-Ferrell type, Larkin-Ovchinnikov type, or a more complicated superposition. That distinction requires the full free-energy landscape at finite momentum.

\section{Crossover of the phase stiffness}

\begin{figure*}[t]
\centering
\includegraphics[width=0.88\textwidth]{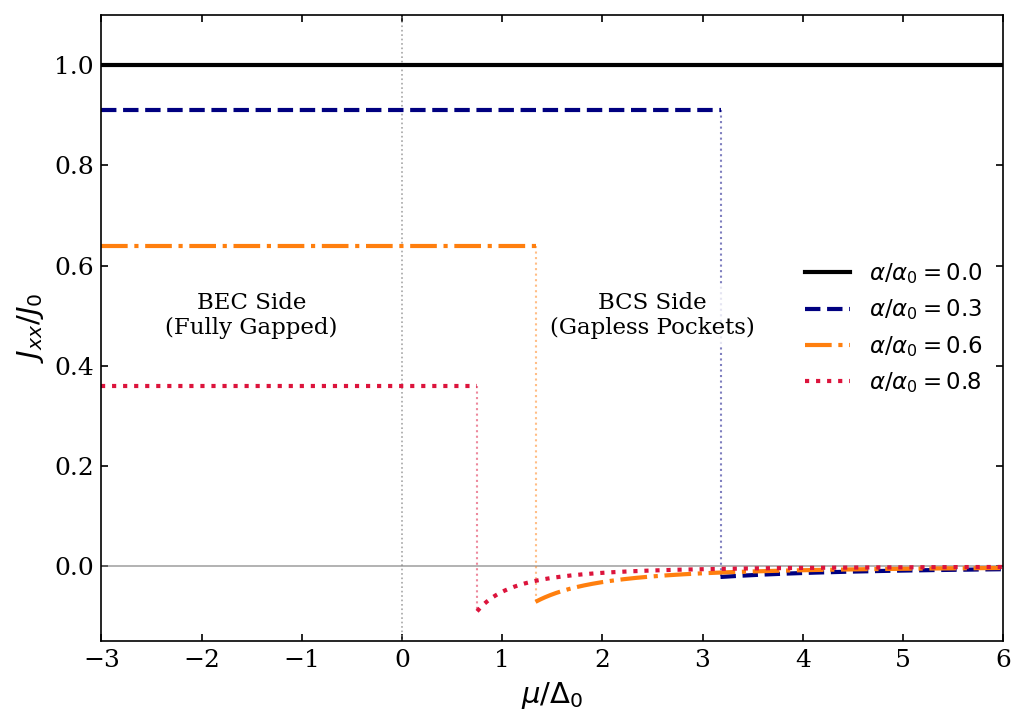}
\caption{Normalized phase stiffness $J_{xx}/J_0$ across the BCS-BEC crossover as a function of the ratio $\mu/\Delta_0$. Curves are shown for representative dimensionless altermagnetic coupling strengths $\alpha/\alpha_0 = 0.0, 0.3, 0.6,$ and $0.8$. Each trajectory is obtained by solving the zero-temperature gap and number equations self-consistently and evaluating the curvature of the finite-momentum free energy. On the BEC side ($\mu/\Delta_0 < 0$), the stiffness remains positive. Instead of collapsing to a single limit, the curves sit on distinct coupling-dependent plateaus, confirming the analytic quadratic renormalization of the effective pair mass, $J_{\rm BEC}/J_0 = 1 - (\alpha/\alpha_0)^2$, derived in the text. Moving into the BCS regime ($\mu/\Delta_0 > 0$), the stiffness remains on these plateaus until the threshold for pocket formation ($\mu/\Delta_0 = \sqrt{1 - a^2}/a$, where $a = \alpha/\alpha_0$) is reached. At this threshold, the stiffness drops discontinuously to a negative value, marking a local instability of the uniform state toward a finite-momentum paired phase.}
\label{fig:stiffness}
\end{figure*}

Figure~\ref{fig:stiffness} summarizes the central result for the phase stiffness along the self-consistent fixed-density BCS-BEC trajectory. The chemical potential and pairing gap entering each point of the curves are obtained from Eqs.~\eqref{eq:gap} and \eqref{eq:number}, and the stiffness is then extracted from the curvature of the finite-momentum free energy defined in Eq.~\eqref{eq:stiffness}.

On the BCS side, the figure shows that the stiffness remains locked to the plateau value $1 - (\alpha/\alpha_0)^2$ as long as the spectrum remains fully gapped. However, when the chemical potential exceeds the threshold $\mu/\Delta_0 = \sqrt{1 - (\alpha/\alpha_0)^2}/(\alpha/\alpha_0)$ and gapless Bogoliubov pockets appear, these low-energy quasiparticles provide an efficient channel for weakening the rigidity of the uniform condensate. At this threshold, the stiffness drops discontinuously to a negative value. This sharp jump marks the loss of local stability of the uniform paired state toward a finite-momentum paired phase.

The BEC side displays a qualitatively different behavior. Once the fermions are bound into compact pairs, the opposite altermagnetic distortions of the two constituents largely cancel in the center-of-mass motion. The stiffness therefore remains positive throughout the deep BEC regime. The curves approach different positive asymptotic values rather than collapsing onto a common plateau. As shown analytically in the next section, this residual separation is the consequence of a quadratic renormalization of the effective pair mass.

The zero of the quadratic stiffness identifies the loss of stability of the uniform state, but it does not by itself determine the spatial structure of the state that replaces it. Distinguishing a single-wavevector Fulde-Ferrell state from a Larkin-Ovchinnikov state or another superposition requires the full free-energy landscape at finite pair momentum. We therefore restrict our focus in Fig.~\ref{fig:stiffness} to the stiffness crossover and do not infer the direction or detailed geometry of the modulated state from the quadratic response alone. While evaluating the free energy at fixed uniform $\mu$ and $\Delta$ is mathematically sufficient to compute the quadratic curvature at the origin (because the first derivatives vanish by symmetry), at finite $\mathbf{q}$ these parameters must be allowed to self-consistently readjust to satisfy the fixed-density constraint. Identifying whether the true global equilibrium is a single-wavevector Fulde-Ferrell phase or a multiple-wavevector Larkin-Ovchinnikov phase requires a full non-linear minimization of the self-consistent finite-momentum free energy, which lies beyond the scope of this work. Our stiffness calculations therefore serve specifically to map the threshold of the local instability of the uniform fully-gapped state.

\section{BCS and BEC limits}

In the weak-coupling BCS regime the chemical potential is positive and the most important fermionic states lie near the effective Fermi surface determined by $\mu$. To distinguish this from the density-defined reference momentum $k_F = \sqrt{4\pi n}$, we define the $\mu$-dependent wavevector as
\begin{equation}
k_\mu=\frac{\sqrt{2m\mu}}{\hbar}.
\label{eq:kmu}
\end{equation}
Let $\phi$ be the polar angle of $\mathbf k$, so that $k_x=k\cos\phi$ and $k_y=k\sin\phi$. Near the Fermi surface, the altermagnetic splitting is approximately
\begin{equation}
J_{\mathbf k}\simeq \frac{2m\alpha\mu}{\hbar^2}\cos(2\phi).
\label{eq:JkBCS}
\end{equation}
This expression makes the directional pair breaking transparent. The mismatch is largest along the principal momentum axes and vanishes along the diagonals. Once gapless pockets form, low-energy quasiparticles can strongly reduce the free-energy curvature and therefore the phase stiffness.

The deep BEC regime admits a complementary interpretation in terms of the center-of-mass motion of a tightly bound pair. We denote the pair wavevector by $\mathbf Q$ and relate it to the phase-gradient wavevector through
\begin{equation}
\mathbf Q=2\mathbf q.
\label{eq:Qdef}
\end{equation}
Consider first motion along the $x$ direction. The two constituents have opposite altermagnetic corrections to their single-particle masses. Completing the square in their combined kinetic energy gives
\begin{equation}
\begin{split}
&(\alpha_0+\alpha)(k_x+Q_x/2)^2+(\alpha_0-\alpha)(-k_x+Q_x/2)^2\\
&=2\alpha_0\left(k_x+\frac{\alpha Q_x}{2\alpha_0}\right)^2
+\frac{\alpha_0}{2}\left[1-\left(\frac{\alpha}{\alpha_0}\right)^2\right]Q_x^2.
\end{split}
\label{eq:pairmassderive}
\end{equation}
Here $Q_x$ is the $x$ component of $\mathbf Q$. The shift of the relative momentum, represented by the first term on the right-hand side, does not change the center-of-mass mass. The coefficient of the second term determines that mass. Repeating the same calculation along $y$ gives the same coefficient, so the composite pair remains isotropic even though its constituents are individually anisotropic.

The resulting effective pair mass is
\begin{equation}
M_B=\frac{2m}{1-(\alpha/\alpha_0)^2}.
\label{eq:MB}
\end{equation}
Here $M_B$ denotes the center-of-mass mass of a tightly bound pair. In the absence of altermagnetism it reduces to the expected value $2m$.

If $J_{\rm BEC}$ denotes the deep-BEC phase stiffness, its value relative to the reference stiffness of Eq.~\eqref{eq:J0} is
\begin{equation}
\frac{J_{\rm BEC}}{J_0}=1-\left(\frac{\alpha}{\alpha_0}\right)^2.
\label{eq:JBEC}
\end{equation}
The absence of a term linear in the altermagnetic strength expresses the cancellation between the two constituents of the pair. The remaining quadratic term shows that the cancellation is strong but not perfect. This distinction is important when the BEC limit is compared with numerical stiffness curves.

Remarkably, the quadratic suppression of the phase stiffness described by Eq.~\eqref{eq:JBEC} is not restricted to the deep BEC limit. At zero temperature, it holds universally across the entire fully gapped phase, which we can now demonstrate explicitly from the finite-momentum free energy. At $T=0$, as long as the quasiparticle spectrum remains fully gapped ($E_{\mathbf{k}} > \vert{}J_{\mathbf{k}}\vert{}$), the Fermi-Dirac occupation terms vanish entirely. Furthermore, due to momentum inversion and reflection symmetries, $\Delta_{\mathbf{q}}$ and $\mu_{\mathbf{q}}$ are even functions of $\mathbf{q}$. Consequently, their first derivatives at $\mathbf{q}=0$ vanish identically and do not contribute to the second-order free-energy curvature. The stiffness is therefore given purely by the explicit momentum dependence of the zero-temperature grand potential:\begin{equation}J \delta_{ij} = \frac{1}{2} \left. \frac{\partial^2}{\partial q_i \partial q_j} \int \frac{d^2\mathbf{k}}{(2\pi)^2} \left( \bar\xi_{\mathbf{k}\mathbf{q}} - E_{\mathbf{k}\mathbf{q}} \right) \right|_{\mathbf{q}=0}.
\end{equation}
Expanding the average dispersion to second order in $\mathbf{q}$ yields $\bar{\xi}_{\mathbf{k}\mathbf{q}} \simeq \xi_{\mathbf{k}} + 2\alpha(k_x q_x - k_y q_y) + \hbar^2 q^2 / (2m)$. Substituting this into the Bogoliubov energy $E_{\mathbf{k}\mathbf{q}} = \sqrt{\bar{\xi}_{\mathbf{k}\mathbf{q}}^2 + \Delta^2}$ and expanding the integrand generates two non-vanishing terms at $O(q^2)$. The first evaluates to $n \hbar^2 q^2 / (2m) = J_0 q^2$, representing the unperturbed stiffness. The second term arises from the square of the anisotropic $O(q)$ correction, which enters with a minus sign from the Taylor expansion of $-E_{\mathbf{k}\mathbf{q}}$:
\begin{equation}
- \int \frac{d^2\mathbf{k}}{(2\pi)^2} \frac{\Delta^2}{2E_{\mathbf{k}}^3} \left[ 2\alpha(k_x q_x - k_y q_y) \right]^2.
\end{equation}
To evaluate this integral, we average over the polar angle and integrate by parts using the derivative identity
\begin{equation}
\frac{d}{dk}\left(1 - \frac{\xi_{\mathbf{k}}}{E_{\mathbf{k}}}\right) = -\frac{\hbar^2 k}{m}\frac{\Delta^2}{E_{\mathbf{k}}^3},
\end{equation}
combined with the zero-temperature number equation $n = \int \frac{d^2\mathbf{k}}{(2\pi)^2} \left(1 - \frac{\xi_{\mathbf{k}}}{E_{\mathbf{k}}}\right)$. This simplifies the second-order integral precisely to $-J_0 (\alpha/\alpha_0)^2 q^2$. Summing these contributions proves that the phase stiffness follows the analytical plateau $J/J_0 = 1 - (\alpha/\alpha_0)^2$ continuously across the crossover, regardless of the sign of the chemical potential. Deviations from this limit occur only when $J_{\mathbf{k}}$ exceeds the energy gap, populating gapless Bogoliubov pockets that provide a new channel for phase fluctuations. By integrating the zero-temperature number equation in the presence of gapless pockets, one finds the relation $2\epsilon_F = (1 + 2/\eta)\mu - R$, where $\eta = \sqrt{1 - (\alpha/\alpha_0)^2}$ and $R = \sqrt{\mu^2 + \Delta_0^2}$. Substituting this into the grand-potential curvature yields the closed-form stiffness on the gapless branch:
\begin{equation}
\frac{J_{xx}}{J_0} = \eta^2 \frac{\mu - \sqrt{\mu^2 + \Delta_0^2}}{2\epsilon_F} < 0.
\label{eq:Jgapless_exact}
\end{equation}
Because $\sqrt{\mu^2 + \Delta_0^2} > \mu$ for any nonzero uniform gap $\Delta_0$, the numerator is negative, proving analytically that the local stiffness remains negative throughout the entire gapless regime.

\section{Conclusions}

We have used a minimal two-dimensional continuum model to compare the effect of altermagnetism on paired fermions across the BCS-BEC crossover. The central physical ingredient is a momentum-dependent spin mismatch that distorts the two spin species in opposite directions. At fixed density, the pairing gap and chemical potential must be determined self-consistently, because the appearance of gapless pockets changes the occupations on the BCS side.

The phase stiffness is most naturally understood as the curvature of the free energy when the pairs are given a small center-of-mass momentum. This viewpoint makes the stability criterion transparent: a positive curvature supports the uniform state, whereas a negative curvature signals an instability toward pairing at finite momentum. The detailed spatial structure of that modulated state requires the full finite-momentum free energy and should not be inferred from the quadratic stiffness alone.

Our analytical derivation demonstrates that at zero temperature, the phase stiffness follows a quadratic suppression relation, $J/J_0 = 1 - (\alpha/\alpha_0)^2$, which holds universally across the entire fully gapped regime of this model. The BEC limit provides a useful physical picture for this plateau: in a tightly bound pair, the opposite altermagnetic distortions of the two fermions cancel at leading order in the pair motion, leaving only this smaller quadratic correction to the composite boson mass. Deviations from this plateau emerge only on the BCS side: precisely when the altermagnetic splitting exceeds the energy gap and opens gapless Bogoliubov pockets, the phase stiffness drops discontinuously below zero. We emphasize that the exactness of this result relies on the zero-temperature continuum limit and a fully gapped spectrum. In real altermagnetic materials, lattice effects beyond the quadratic dispersion, finite temperatures, and complex orbital-dependent interactions will inevitably introduce non-universal corrections. Furthermore, while the full BCS-BEC crossover is highly tunable in ultracold atomic gases, tuning the interaction strength $\epsilon_B/\epsilon_F$ over such a broad range in solid-state altermagnets remains a significant experimental challenge. Therefore, rather than a prediction for a specific material, our continuum limits provide a theoretical baseline on the stability of uniform singlet pairing against directional spin-splitting.
\appendix
\section{Microscopic motivation for the continuum splitting}
\label{app:lattice}

The continuum splitting of Eq.~\eqref{eq:Jk} can be motivated by a simple square-lattice model with two interpenetrating sublattices, labeled $A$ and $B$. We set the lattice constant equal to one, so the components of $\mathbf k$ in this appendix are dimensionless crystal momenta. Let $t'$ denote the average hopping amplitude within a sublattice and let $t_d$ measure the difference between hopping along the two principal directions. The isotropic part of the tight-binding dispersion is
\begin{equation}
\epsilon_0(\mathbf k)=-2t'\left(\cos k_x+\cos k_y\right).
\label{eq:epsilon0lattice}
\end{equation}
The directional part is
\begin{equation}
\gamma(\mathbf k)=-2t_d\left(\cos k_x-\cos k_y\right).
\label{eq:gammak}
\end{equation}
The function $\gamma(\mathbf k)$ distinguishes the $x$ and $y$ directions. We also introduce an inter-sublattice hybridization amplitude $V_0$ and an exchange field of magnitude $h$ that has opposite sign on the two sublattices.

For a spin label $\sigma=+1$ for spin up and $\sigma=-1$ for spin down, the mean-field Hamiltonian is
\begin{equation}
H_{\rm MF}(\mathbf k,\sigma)=
\begin{pmatrix}
\epsilon_0(\mathbf k)+\gamma(\mathbf k)-\sigma h & V_0\\
V_0 & \epsilon_0(\mathbf k)-\gamma(\mathbf k)+\sigma h
\end{pmatrix}.
\label{eq:latticeH}
\end{equation}
Its two bands are
\begin{equation}
E_{\lambda}(\mathbf k,\sigma)=\epsilon_0(\mathbf k)+\lambda\sqrt{V_0^2+[\gamma(\mathbf k)-\sigma h]^2}.
\label{eq:latticebands}
\end{equation}
Here $\lambda=+1$ labels the upper band and $\lambda=-1$ labels the lower band. The momentum-dependent anisotropy $\gamma(\mathbf k)$ distinguishes the two spatial directions, while the field $h$ distinguishes the two spin projections in opposite ways on the two sublattices. Together they generate the momentum-dependent spin splitting characteristic of the altermagnetic state.

Expanding the lower band near the center of the Brillouin zone gives
\begin{equation}
\gamma(\mathbf k)\simeq t_d(k_x^2-k_y^2).
\label{eq:gammacont}
\end{equation}
At Hartree-Fock level, an on-site Hubbard repulsion $U$ and a staggered sublattice magnetization $M$ generate the exchange field $h=UM$. The resulting continuum altermagnetic coupling is
\begin{equation}
\alpha=\frac{h\,t_d}{\sqrt{V_0^2+h^2}}.
\label{eq:alphamicroscopic}
\end{equation}
Thus the simple continuum parameter $\alpha$ can be viewed as the low-energy remnant of the combined hopping anisotropy and staggered magnetic order. The appendix is intended only as a microscopic motivation for the continuum form used in the main text; the crossover analysis itself does not require the lattice model.

\begin{acknowledgments}
L.S. is partially supported by the Project ``Frontiere Quantistiche'' (Dipartimenti di Eccellenza) of the Italian Ministry of University and Research and by ``Iniziativa Specifica Quantum'' of INFN. We are grateful to an anonymous referee for rigorous review and for providing the analytical derivation of the phase stiffness.
\end{acknowledgments}


\begin{thebibliography}{99}

\bibitem{hayami2019}
S. Hayami, Y. Yanagi, and H. Kusunose,
Momentum-Dependent Spin Splitting by Collinear Antiferromagnetic Ordering,
J. Phys. Soc. Jpn. \textbf{88}, 123702 (2019).

\bibitem{hayami2026}
T. Sato and S. Hayami,
Quantum theory of magnetic octupole in periodic crystals and application to d-wave altermagnets,
npj Quantum Mater. \textbf{11}, 32 (2026).

\bibitem{smejkal2022beyond}
L. Smejkal, J. Sinova, and T. Jungwirth,
Beyond Conventional Ferromagnetism and Antiferromagnetism: A Phase with Nonrelativistic Spin and Crystal Rotation Symmetry,
Phys. Rev. X \textbf{12}, 031042 (2022).

\bibitem{smejkal2022landscape}
L. Smejkal, J. Sinova, and T. Jungwirth,
Emerging Research Landscape of Altermagnetism,
Phys. Rev. X \textbf{12}, 040501 (2022).

\bibitem{fulde1964}
P. Fulde and R. A. Ferrell,
Superconductivity in a Strong Spin-Exchange Field,
Phys. Rev. \textbf{135}, A550 (1964).

\bibitem{larkin1965}
A. I. Larkin and Yu. N. Ovchinnikov,
Nonuniform state of superconductors,
Sov. Phys. JETP \textbf{20}, 762 (1965).

\bibitem{casalbuoni2004}
R. Casalbuoni and G. Nardulli,
Inhomogeneous superconductivity in condensed matter and QCD,
Rev. Mod. Phys. \textbf{76}, 263 (2004).

\bibitem{chakraborty2024finiteq}
D. Chakraborty and A. M. Black-Schaffer,
Zero-field finite-momentum and field-induced superconductivity in altermagnets,
Phys. Rev. B \textbf{110}, L060508 (2024).

\bibitem{banerjee2024diode}
S. Banerjee and M. S. Scheurer,
Altermagnetic superconducting diode effect,
Phys. Rev. B \textbf{110}, 024503 (2024).

\bibitem{fu2025}
P.-H. Fu, Q. Lv, Y. Xu, J. Cayao, J.-F. Liu, and X.-L. Yu,
All-electrically controlled spintronics in altermagnetic heterostructures,
npj Quantum Mater. \textbf{10}, 111 (2025).

\bibitem{fu2026}
P.-H. Fu, S. Mondal, J.-F. Liu, Y. Tanaka, and J. Cayao,
Floquet Engineering Spin Triplet States in Unconventional Magnets,
Phys. Rev. Lett. \textbf{136}, 066703 (2026).

\bibitem{cayao2025}
Y. Fukaya, B. Lu, K. Yada, Y. Tanaka, and J. Cayao,
Superconducting phenomena in systems with unconventional magnets,
J. Phys.: Condens. Matter \textbf{37}, 313003 (2025).

\bibitem{liu2026altermagnetic}
X.-J. Liu and H. Hu,
Altermagnetism-driven FFLO superconductivity in finite-filling 2D lattices,
AAPPS Bulletin \textbf{36}, 6 (2026).

\bibitem{hong2025}
S. Hong, M. J. Park, and K.-M. Kim,
Unconventional p-wave and finite-momentum superconductivity induced by altermagnetism through the formation of Bogoliubov Fermi surface,
Phys. Rev. B \textbf{111}, 054501 (2025).

\bibitem{delasheras2025}
R. de las Heras, T. Kokkeler, S. Ili{\'c}, I. V. Tokatly, and F. S. Bergeret,
Interplay between Superconductivity and Altermagnetism in Disordered Materials and Heterostructures,
arXiv:2512.04819 (2025).

\bibitem{liu2026fflo}
Z. Liu, H. Hu, X. J. Liu, 
Fulde-Ferrell-Larkin-Ovchinnikov states and topological Bogoliubov Fermi surfaces in altermagnets: An analytical study,
Phys. Rev. B \textbf{113}, 024518 (2026).

\bibitem{randeria1989}
M. Randeria, J.-M. Duan, and L.-Y. Shieh,
Bound states, Cooper pairing, and Bose condensation in two dimensions,
Phys. Rev. Lett. \textbf{62}, 981 (1989).

\bibitem{randeria1990}
M. Randeria, J.-M. Duan, and L.-Y. Shieh,
Superconductivity in a two-dimensional Fermi gas: Evolution from Cooper pairing to Bose condensation,
Phys. Rev. B \textbf{41}, 327 (1990).

\bibitem{chakraborty2025constraints}
D. Chakraborty and A. M. Black-Schaffer,
Constraints on superconducting pairing in altermagnets,
Phys. Rev. B \textbf{112}, 014516 (2025).

\bibitem{Wang2014}
P.-J. Wang and J. Zhang,
Spin-orbit coupling in Bose-Einstein condensate and degenerate Fermi gases,
Frontiers of Physics \textbf{9}, 598--612 (2014).

\bibitem{Cheuk2012}
L. W. Cheuk, A. T. Sommer, Z. Hadzibabic, T. Yefsah, W. S. Bakr, and M. W. Zwierlein,
Spin-Injection Spectroscopy of a Spin-Orbit Coupled Fermi Gas,
Phys. Rev. Lett. \textbf{109}, 095302 (2012).

\bibitem{Zhai2015}
H. Zhai,
Degenerate quantum gases with spin-orbit coupling: a review,
Rep. Prog. Phys. \textbf{78}, 026001 (2015).

\bibitem{DellAnna2011}
L. Dell'Anna, G. Mazzarella, and L. Salasnich,
Condensate fraction of a resonant Fermi gas with spin-orbit coupling in three and two dimensions,
Phys. Rev. A \textbf{84}, 033633 (2011).

\bibitem{DellAnna2012}
L. Dell'Anna, G. Mazzarella, and L. Salasnich,
Tuning Rashba and Dresselhaus spin-orbit couplings: Effects on singlet and triplet condensation with Fermi atoms,
Phys. Rev. A \textbf{86}, 053632 (2012).

\bibitem{maeda2025}
K. Maeda, Y. Fukaya, K. Yada, B. Lu, Y. Tanaka, and J. Cayao, 
Classification of pair symmetries in superconductors with unconventional magnetism,
Phys. Rev. B \textbf{111}, 144508 (2025).

\bibitem{yang2026}
F. Yang, G.-D. Zhao, B. Yan, and L.-Q. Chen,
Emergent spin-resolved electronic charge density waves and pseudogap phenomena from strong 3-wave altermagnetism,
arXiv:2602.04511 (2026).

\bibitem{choi2026exploring}
I. H. Choi, S. G. Jeong, B. Jalan, and J. S. Lee,
Exploring altermagnetism in RuO2: from conflicting experiments to emerging consensus,
Nano Convergence \textbf{13}, 1 (2026).

\bibitem{jeong2026altermagnetic}
S. G. Jeong \textit{et al.},
Altermagnetic polar metallic phase in ultrathin epitaxially strained RuO2 films,
Proc. Natl. Acad. Sci. U.S.A. \textbf{123}, e2526641123 (2026).

\bibitem{lee2026strain}
S. Lee, S. G. Jeong, J.-P. Wang, B. Jalan, and T. Low,
Strain-Driven Altermagnetic Spin-Splitting Effect in RuO2,
Nano Lett. \textbf{26}, 8110 (2026).

\bibitem{shabani2016two}
J. Shabani \textit{et al.},
Two-dimensional epitaxial superconductor-semiconductor heterostructures: A platform for topological superconducting networks,
Phys. Rev. B \textbf{93}, 155402 (2016).

\bibitem{kjaergaard2016quantized}
M. Kjaergaard \textit{et al.},
Quantized conductance doubling and hard gap in a two-dimensional semiconductor-superconductor heterostructure,
Nat. Commun. \textbf{7}, 12841 (2016).

\bibitem{khodas2026tuning}
M. Khodas, S. Mu, I. I. Mazin, and K. D. Belashchenko,
Tuning of altermagnetism by strain,
Phys. Rev. B \textbf{113}, 104422 (2026).

\bibitem{zhang2026tunable}
T. Zhang, L. Yuan, J. M. Rondinelli, H. A. Fertig, and S. Zhang,
Tunable Hidden Altermagnetic Spin Splitting in Layered Ruddlesden–Popper Oxides,
Nano Lett. \textbf{26}, 2778 (2026).

\bibitem{wang2026strain}
J. Wang, W. Zhang, Y. Liu, J. Hu, Z. Zhang, R. Xiong, and Z. Lu,
Strain-engineered modulation of non-relativistic altermagnetic spin splitting in rutile RuO2,
Mater. Today Phys., 102081 (2026).

\end{thebibliography}
\end{document}